\documentclass[10pt,twocolumn,letterpaper]{article}
\usepackage{booktabs,makecell,siunitx,adjustbox,xcolor}
\renewcommand{\arraystretch}{1.12}
\newcommand{\upg}[1]{\textcolor{green!60!black}{\scriptsize$\uparrow$~#1}}
\newcommand{\downg}[1]{\textcolor{green!60!black}{\scriptsize$\downarrow$~#1}}
\newcommand{\upr}[1]{\textcolor{red!70!black}{\scriptsize$\uparrow$~#1}}

\usepackage{wacv}              

 \usepackage{multirow} 

\definecolor{wacvblue}{rgb}{0.21,0.49,0.74}
\usepackage[pagebackref,breaklinks,colorlinks,allcolors=wacvblue]{hyperref}

\def\wacvPaperID{*****} 
\def\confName{WACV}
\def\confYear{2026}

\title{SymNet: A Multi-Task Network for Joint Radio Map Reconstruction and Transmitter Localization}

\author{Lyuzhou Ye\\
University of North Texas\\
1155 Union Circle\\
Denton, Texas\\
{\tt\small lyuzhouye@my.unt.edu}
\and
Thanh Dat Le\\
University of North Texas\\
1155 Union Circle\\
Denton, Texas\\
{\tt\small ThanhLe5@my.unt.edu}
\and
Yan Huang\\
University of North Texas\\
1155 Union Circle\\
Denton, Texas\\
{\tt\small Yan.Huang@unt.edu}
}
\begin{document}
\maketitle
\begin{abstract}
Accurately predicting directional radio maps is essential for wireless applications, yet prior approaches primarily focus on omnidirectional signals and typically treat transmitter localization and signal map reconstruction as separate tasks. In omnidirectional settings, predicting the maximum signal location often coincides with the transmitter position, which limits the need for explicit joint modeling. However, in directional propagation where angular effects, reflections, and building occlusions play critical roles, this assumption no longer holds. To address this gap, we propose SymNet, a unified framework that jointly predicts directional radio maps and transmitter locations from sparse signal measurements. SymNet incorporates a prediction head for transmitter localization alongside radio map reconstruction, enabling simultaneous learning of both tasks. This joint formulation leverages their complementary information and leads to consistent improvements over treating them separately. Experiments on challenging directional scenarios demonstrate that SymNet outperforms state-of-the-art baselines, achieving superior accuracy in both radio map reconstruction and transmitter localization.
\end{abstract}
\vspace{-2em}    
\section{Introduction}
\label{sec:intro}
Wireless systems increasingly rely on directional propagation (e.g., in mmWave and beamforming-based 5G), where received signal strength is highly sensitive to angles, occlusions, and reflections. A fundamental challenge in this setting is to reconstruct accurate directional radio maps from sparse measurements and to determine the transmitter location simultaneously. Both tasks are critical for applications such as network planning, localization, and resource management, yet they are often studied in isolation.
\vspace{-0.2em}

On the one hand, recent learning-based approaches have advanced radio map prediction by leveraging convolutional or U-Net-like architectures \cite{Levie2019RadioUNetFR, SkipNet, GAN, DCNet}. These models focus on reconstructing omnidirectional signal distributions and achieve strong performance compared to classical interpolation methods, but typically assume that the transmitter position can be directly inferred from the heatmap maximum. Meanwhile, localization-focused methods such as LocNet \cite{LocNet}, TLDL \cite{tldl}, and DSLoc \cite{dsloc} employ U-Net-style encoder–decoder networks (or HRNet in DSLoc) to directly regress transmitter heatmaps and extract positions. While effective in omnidirectional or simplified propagation scenarios, these methods do not explicitly address the broader radio map reconstruction problem.
\vspace{-0.2em}

In directional settings, the strongest signal region often deviates from the transmitter location due to angular effects and multi-path propagation. This breaks the implicit alignment between the two tasks, making radio map reconstruction and transmitter localization not only distinct but also potentially complementary. Despite this, prior work in both areas has either reconstructed maps without explicit localization or localized transmitters without modeling the full signal distribution. Our contributions are as follows:

\begin{itemize}
    \item We identify and formulate the joint prediction problem of directional radio maps and transmitter localization, a task not explicitly studied in omnidirectional-focused studies.
    \item We present SymNet, a novel architecture that couples heatmap reconstruction with transmitter prediction through a shared backbone and task-specific heads.
    \item Through experiments in challenging directional scenarios, we show that SymNet achieves consistent gains in both tasks compared to state-of-the-art methods: it reduces localization error (meters) by an average of 34.36\% relative to DSLoc across 20–100 samples, lowers \textbf{RMSE} by 4.08\% compared with ViT-RefineNet, and improves \textbf{SSIM} by +4.70\% on average. Moreover, it surpasses the omnidirectional SOTA DC-Net by 15\% in \textbf{RMSE}, highlighting its advantage under angular-sensitive and multipath-dominated conditions.
\end{itemize}

\section{Related work}
\label{sec:relatedwork}
\subsection{Methods for Radio Map Generation}
Classical approaches for constructing radio maps can be broadly divided into parametric and non-parametric methods. Parametric techniques, such as compressed sensing \cite{Compressed_sensing} and dictionary learning \cite{dictionary}, assume detailed knowledge of transmitter characteristics. Non-parametric methods, such as Ordinary Kriging \cite{Oridinary_Kriging, interpolation} and Radial Basis Function interpolation \cite{RadialBasic}, rely on hand-crafted models of signal propagation. Both categories face substantial limitations in complex urban environments, where multipath effects and severe attenuation often invalidate their underlying assumptions.

Recent work on deep learning–based radio map generation can be divided into two categories: those assuming known transmitter locations and those assuming unknown transmitter locations. When the transmitter position is known, models such as RadioUNet \cite{Levie2019RadioUNetFR} and its extensions \cite{UNETSI, PPNET, PMNET} learn robust propagation patterns by combining location priors with environmental data. More lightweight designs, including GAN-based models \cite{BASICGAN, PLGAN} and Transformer-augmented autoencoders \cite{RADIONET}, have further improved efficiency and map realism.

When the transmitter position is unknown, the task becomes significantly more challenging, as models must infer radio maps solely from sparse receiver measurements without prior knowledge of the source. This setting has been the focus of much recent research.

Teganya et al. \cite{DEEPAE} first demonstrated that a U-Net could generate accurate omnidirectional radio maps using only sparse samples and a coarse environment map. Building on this idea, SkipNet \cite{SkipNet} introduced lightweight skip connections to improve stability, while RadioUNet-S \cite{Levie2019RadioUNetFR} explored deeper variants to increase representational capacity. Generative approaches such as GAN-CRME \cite{CGAN} streamlined the U-Net generator, utilizing adversarial training to enhance realism with reduced computational cost. Enes et al. \cite{twinUnets} further advanced this line by employing twin networks to jointly predict signal power and uncertainty, improving robustness in highly variable environments.

Despite these advances, convolutional designs fundamentally struggle to capture long-range dependencies, which are critical for directional propagation. To overcome this, later methods have emphasized multi-scale context aggregation. ACT-GAN \cite{ACTGAN} and DC-Net \cite{DCNet} mimicked Transformer-like receptive fields through pyramid pooling and dilated convolutions, partially alleviating locality constraints, though these approximations may struggle to capture precise angular sensitivities or long-range interactions.

Most recently, ViT-RefineNet \cite{ViTRefineNet} introduced a hybrid design that combines a Vision Transformer backbone with U-Net-like refinement modules, effectively balancing global dependency modeling with fine-grained detail recovery. Although achieving strong performance in the prediction of the directional radio map, this model, similar to prior approaches, it focused exclusively on heatmap reconstruction and did not address transmitter localization as a distinct and complementary task.

\subsection{Methods for Transmitter Localization}
Traditional transmitter localization methods, such as time of arrival (TOA), time difference of arrival (TDOA), and angle of arrival (AOA) \cite{gezici2008survey, gustafsson2005mobile}, require complex hardware support and are difficult to deploy at scale. RSS-based localization is easier to implement \cite{832252,patwari2005locating} but is highly sensitive to non-line-of-sight propagation and typically depends on propagation models that fail in dense urban environments \cite{haeberlen2004practical}.

With the rise of deep learning, researchers have explored data-driven localization frameworks that learn propagation characteristics directly from measurements. Coordinate-based approaches predict the transmitter’s $(x,y)$ location explicitly. Zhang et al \cite{zhang}. pioneered the use of deep learning for transmitter localization, combining an MLP feature extractor with an HMM to directly predict transmitter coordinates. DeepTxFinder \cite{zubow2020deeptxfinder} extended this direction by mapping sparse RSS readings into grid cells and using a CNN–MLP design to estimate both the number and positions of transmitters, showing good performance in dense sampling scenarios. Building on this line, Wang et al. \cite{wang} proposed MT-GCNN, a multi-task gated CNN that jointly performs containment cell classification and distance regression, improving robustness under varying sensor deployments. While effective, these coordinate-based methods often face generalization challenges when sensor layouts change or sampling density is low.

More recently, heatmap-based approaches have become dominant. DeepMTL \cite{deepmtl} reframed localization as object detection, adapting YOLOv3 to produce transmitter probability maps. TL;DL \cite{tldl} predicts transmitter heatmaps with a U-Net and then thresholds/suppresses peaks to extract transmitter coordinates from the output, which is capable of localizing multiple transmitters. LocNet \cite{LocNet} further demonstrated that a lightweight U-Net variant can achieve competitive accuracy with orders of magnitude fewer parameters. Most recently, DSLoc \cite{dsloc} addressed the extreme sparsity challenge by introducing bias correction and sparse expansion preprocessing, adopting HRNet as the backbone to preserve high-resolution features, and replacing argmax with centroid regression to improve localization precision.

While these deep learning approaches have achieved promising results, they are primarily localization-focused solutions and generally do not attempt to reconstruct full radio maps and are largely developed under omnidirectional assumptions, where the transmitter location aligns with the signal maximum. However, in directional propagation, this assumption no longer holds, and localization and radio map prediction become distinct but complementary tasks. This gap motivates our proposed SymNet, which unifies both tasks in a joint framework.
\section{Preliminary}
\subsection{Problem Definition}
Given a region of interest (ROI) that is divided into a two-dimensional grid of $H \times W$, where $H$ and $W$ are the height and width of the grid, respectively, with ($H, W \in \mathbb{Z}^+$). A single transmitter, equipped with a directional antenna whose orientation (0–360 degrees) is randomly assigned in each scenario, is located at $(T_x, T_y)$ within this grid. A subset of pixels in the area is randomly sampled. Our goal is to develop a model to predict the radio strength at each pixel in the given region as well as to predict the position of the transmitter.
\subsection{Grid-based Representation}
We use two matrices to represent the environment.
The first matrix is the building map $\mathbf{E} = (b_{i,j})$, where
\[
b_{i,j} =
\begin{cases}
-1, & \text{if $(i,j)$ is a building pixel}, \\[6pt]
1,  & \text{if $(i,j)$ is a sampled pixel}, \\[6pt]
0,  & \text{if $(i,j)$ is a non-building non-sampled pixel}.
\end{cases}
\]

The second matrix is the signal map $\mathbf{R} = (r_{i,j})$, defined as
\[
r_{i,j} =
\begin{cases}
s_{i,j}, & \text{if $(i,j)$ is sampled (with reading $s_{i,j}$)}, \\[6pt]
0,       & \text{otherwise}.
\end{cases}
\]

\section{Proposed Model Framework}
Our model architecture, illustrated in Figure \ref{PA} (Top), is composed of several modules connected through a sequential data flow.We concatenate the output of each module with the original environment information at the fusion points to ensure that the model utilizes environmental information.

The data flow begins with a SkipNet with attention gates that enhances local features. Its output, after concatenation with the original environment map, is fed into two parallel paths to capture the global contextual information, one for heatmap generation and the other for transmitter location prediction. After patchification and embedding, the data passes through six self-attention encoding blocks of the Vision Transformer \cite{VIT}, the outputs are then fused by one cross-attention module, and added back to the original outputs. Then the two separated fused outputs are fed into two separate self-attention encoding blocks again,  decoded by two lightweight decoders, concatenated with each other and then the environment map again, and passed to two extra SkipNet modules for final output. 

This design is inspired by the intuitive relationship between the two tasks: the signal heatmap provides structural cues that narrow the plausible transmitter region, while the estimated location offers global guidance that helps refine the heatmap.
\subsection{Model Input}
\begin{figure*}[t]
    \centering
    \includegraphics[width=1\textwidth]{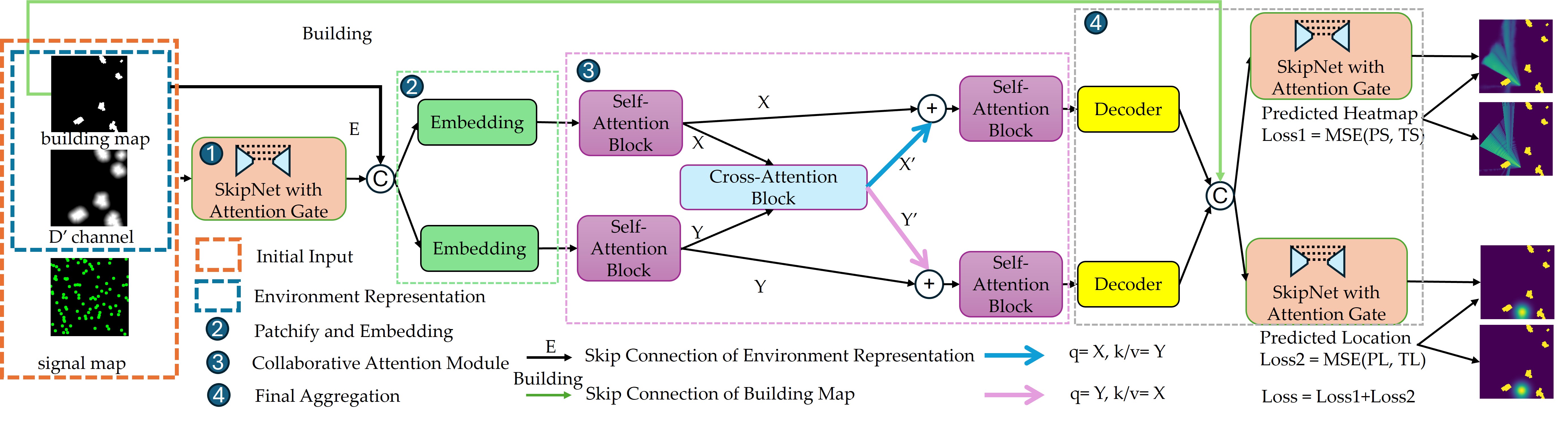}
    \caption{Proposed Framework for Radio Map Prediction. The outputs after cross-task fusion are denoted as X' and Y'. All the LayerNorm normalizations are replaced with Dynamic-Tanh. Modules at different positions do not share parameters.}
    \vspace{-2em}
    \label{PA}
\end{figure*}

For the input of our model in Figure \ref{PA} (Left), in addition to the building map and signal map, we compute the distance from each pixel to its nearest building pixel and construct an additional channel called 
\textbf{Distance from Nearest Building (DNB)}, denoted as 
$\mathbf{D} = (d_{i,j})$, where
\[
d_{i,j} = \min_{(p,q) \in \mathcal{B}} \sqrt{(i-p)^2 + (j-q)^2},
\]
and $\mathcal{B} = \{(p,q) \mid b_{p,q} = -1\}$ denotes the set of all building pixels.
We normalize this data as follows:
\[
d'_{i,j} =
\begin{cases}
1 - \dfrac{d_{i,j}}{d}, & 0 \leq d_{i,j} \leq d, \\[8pt]
0, & d_{i,j} > d.
\end{cases}
\]
denote the result as $D'$ and and include it as one of the model’s input channels.
This design encourages the model to focus on building edge regions, thus improving its ability to learn signal reflection patterns.
\subsection{\textcircled{1} SkipNet with Attention Gates}
SkipNet \cite{SkipNet}, based on the U-Net architecture, is an encoder-decoder network with skip connections. The encoder progressively downsamples the input to extract hierarchical features, while the decoder upsamples and receives the hierarchical features from the encoder through skip connections to reconstruct the spatial resolution for final prediction. LocNet \cite{LocNet} took a similar architecture and validated the effectiveness in the localization task as well. In ViT-RefineNet \cite{ViTRefineNet}, they incorporated attention gates into the SkipNet module, reducing the RMSE of the final prediction. 
\vspace{-1em}
\subsection{\textcircled{2} Patchify and Embedding}
This module was proposed in original ViT \cite{VIT}, which divides an image into $\frac{256}{P}\times \frac{256}{P}$ patches of equal size $P\times P$. 
After patch embedding, we add a fixed 2D positional encoding over the patch grid.
Let $(x,y)$ be the patch coordinates on a $\frac{256}{P}\times \frac{256}{P}$ grid and $d$ the embedding
dimension. The subscript $[2k]$ (or $[2k+1]$) indexes the $(2k)$-th (or $(2k\!+\!1)$-th) element in the $d$-dimensional vector. We allocate half of the channels to $x$ and the other half to $y$:
\[
\scalebox{0.8}{$
\begin{aligned}
\text{for } k=0,\dots,\tfrac{d_x}{2}-1:
&PE_x[2k]=\sin\!\Big(\tfrac{x}{10000^{\,2k/d_x}}\Big),\\\;
&PE_x[2k\!+\!1]=\cos\!\Big(\tfrac{x}{10000^{\,2k/d_x}}\Big),\\
\text{for } k=0,\dots,\tfrac{d_y}{2}-1:
&PE_y[2k]=\sin\!\Big(\tfrac{y}{10000^{\,2k/d_y}}\Big),\\\;
&PE_y[2k\!+\!1]=\cos\!\Big(\tfrac{y}{10000^{\,2k/d_y}}\Big),\\
\text{and}\quad &\mathrm{PE}(x,y)=\big[\,PE_x\;;\;PE_y\,\big]\in\mathbb{R}^{d},
\end{aligned}$}
\]
where $d_x=d_y=\tfrac{d}{2}$. (Note that $d\equiv 0 \text{ mod 4}$.)

\subsection{\textcircled{3} Collaborative Attention Module}

The attention encoder blocks operate on embedded image patches. 
Let $X \in \mathbb{R}^{n \times d}$ denote the input sequence of embedded patches, 
where $n$ is the number of patches and $d$ is the embedding dimension. 
Before entering the attention layers, the sequence is normalized by a learnable 
dynamic-tanh function \cite{tanh}, defined as:
\begin{equation}
    \text{DT}(X) = W \odot \tanh(\alpha X) + B,
\end{equation}
where $\alpha$ is a learnable scalar and 
$\tanh(x) = \frac{e^{2x}-1}{e^{2x}+1}$ is the hyperbolic tangent function. 

\subsubsection{Self-Attention }
The normalized patches are processed by a Multi-Head Self-Attention (MHSA) module, 
which models contextual dependencies within the input sequence. 
The standard query–key–value formulation is as follows:
\[
Q = XW^Q, \quad K = XW^K, \quad V = XW^V,
\]
with $W^Q, W^K, W^V \in \mathbb{R}^{d \times d_k}$. 
The scaled dot-product attention is:
\[
\text{Attention}(Q, K, V) = \text{softmax}\!\left(\tfrac{QK^\top}{\sqrt{d_k}}\right)V.
\]
MHSA uses multiple parallel heads and concatenates their outputs:
\[
\text{MultiHead}(X) = \text{Concat}(\text{head}_1, \ldots, \text{head}_h)W^O,
\]
where each $\text{head}_i = \text{Attention}(Q_i, K_i, V_i)$ and $W^O \in \mathbb{R}^{hd_k \times d}$. 

\subsubsection{Cross-Attention}
To enable interaction between the two prediction tasks, radio map reconstruction
and transmitter localization, we introduce a task-level cross-attention module. 
Let $X$ and $Y$ denote the hidden representations for the radio map 
and localization heads, respectively. 
Cross-attention allows one task to use the features of the other as context:
\[
Q_X = XW^Q, \quad K_Y = Y W^K, \quad V_Y = Y W^V,
\]
so that the radio map head can attend to localization cues. 
Symmetrically, localization features attend to radio map representations: 
\[
Q_Y = Y W^Q, \quad K_X = X W^K, \quad V_X = X W^V.
\]
The attention outputs from all heads are also concatenated and linearly projected, 
and the final output is obtained by a residual connection followed by Dynamic-Tanh:
\begin{align}
X' &= \operatorname{DT}\!\left(X + \text{MultiHead}(Q_X,K_Y,V_Y)\right), \\
Y' &= \operatorname{DT}\!\left(Y + \text{MultiHead}(Q_Y,K_X,V_X)\right).
\end{align}
This bi-directional interaction encourages complementary learning: 
the radio map task captures fine-grained propagation patterns, whereas the localization task enforces spatial consistency around the transmitter position.
The output of the cross-attention module will go through another six self-attention blocks for further refinement.
\subsubsection{Residual and Feed-Forward in Self/Cross Attention Blocks}
Both self-attention and cross-attention outputs are followed by residual connections, 
Dynamic-Tanh normalization, and a feed-forward multilayer perceptron (MLP) block 
with nonlinear activation. Each sub-block uses an additional residual connection. 
Following \cite{tanh}, we replace traditional normalization with the dynamic-tanh 
regularization to improve stability and representation capacity. 

\subsection{\textcircled{4} Final Aggregation}
After the collaborative attention module, following common practice in Vision Transformer–based dense prediction architectures, we employ a lightweight decoder (we denote this module as decoder in Figure \ref{P2}) to restore spatial structure from the refined patch embeddings: each resulting patch embedding is linearly projected back to the pixel-level representation. These projected embeddings are then reshaped and rearranged into a $N \times N$ 2D grid to align with the original spatial layout, concatenated with the original building map, and fed to two different SkipNets for the final heatmap and localization prediction.

\section{Experiment Design and Result}
In this section, we conduct extensive experiments to evaluate the
performance of our method.
In the directional signal scenario, it is common to sample zero-signal pixels. Therefore, we evaluate the impact of the positive ratio (\textbf{pr}), defined as \[\textbf{pr}=\frac{\text{\# of sampled pixels with nonzero signal strength}}{\text{\# of sampled pixels}}\times 100\%.\]
We seek answers to the following research questions, comparing with both radio map prediction methods and transmitter localization methods:
\begin{itemize}
    \item {\bf RQ1:} How does our model perform compared to the baseline models as the number of sampled points increases?
    \item {\bf RQ2:} How does our model perform compared to the state-of-the-art models with different positive ratios.
    \item {\bf RQ3:} How does our model perform compared to the state-of-the-art models over varying average distances between the sampled pixels and the transmitter?
    \item {\bf RQ4:} How do the designs and modules in our model affect performance?
\end{itemize}
See Table \ref{t2} for the factors we plan to test.
\begin{table}[tbp]
\vspace{-0.5em}  
\caption{Definitions and value ranges of factors used to evaluate model performance.}
\centering
\begin{adjustbox}{width=\columnwidth}
\begin{tabular}{|l|ll|}
\hline
\textbf{Factors} & \textbf{Description} & \textbf{Possible Values} \\
\hline
\#samples ($\mathbf{\Omega}$) & Number of sampled points. & 20, 40, 60, 80, 100 \\
Positive Ratio ($\mathbf{pr}$) & Proportion of nonzero signals. & Varies (0\%–100\%) \\
$d_{\text{mean}}$& Average distance from sampled pixels to the transmitter&[0, $256\sqrt{2}$]\\
\hline

\end{tabular}
\end{adjustbox}
\vspace{-2em}
\label{t2}

\end{table}
\subsection{Comparison Models}
For the radio map generation task, we compare our proposed method with the following state-of-the-art models:
\begin{itemize}
    \item \textbf{RadioUNet} \cite{Levie2019RadioUNetFR}: A standard U-Net architecture characterized by its use of deep convolutional filters.
    \item \textbf{SkipNet} \cite{SkipNet}: A compact U-Net variant, similar to RadioUNet, which utilizes a constant number of filters across its corresponding encoder and decoder layers.
    \item \textbf{GAN-CRME} \cite{CGAN}: A Conditional Generative Adversarial Network (c-GAN) designed specifically for lightweight and efficient generation of radio maps.
    \item \textbf{DC-Net} \cite{DCNet}: A model featuring cascaded U-Nets. It integrates Pyramid Spatial Pooling at each downsampling layer of the encoder to improve feature extraction, while the second U-Net in the cascade works to refine and correct the initial output.
    \item \textbf{ViT-RefineNet} \cite{ViTRefineNet}: A model using ViT as a backbone and SkipNet for multistep refinement, the state-of-the-art model for the directional signal dataset.
\end{itemize}

For the transmitter prediction task, we compare our proposed method with the following state-of-the-art models:
\begin{itemize}
    \item \textbf{LocNet} \cite{LocNet}: A lightweight U-Net/SkipNet-style encoder–decoder that predicts transmitter heatmaps; uses \textit{Focal Loss} to address class imbalance.
    \item \textbf{TL;DL} \cite{tldl}: A U-Net-based image-to-image model with a large receptive field that maps sparse/noisy RSSI grids to transmitter heatmaps, followed by thresholding and local-maximum suppression to extract coordinates.
    \item \textbf{DSLoc} \cite{dsloc}: Addresses extreme sparsity via bias-correction and sparse-expansion preprocessing; adopts \textit{HRNet} to preserve high-resolution features and replaces argmax with centroid regression for finer localization.
\end{itemize}
\begin{table}[ht]
\caption{Hyperparameter settings.}
\centering
\renewcommand{\arraystretch}{0.5} 
\begin{tabular}{ll}
\toprule
\textbf{Hyperparameter} & \textbf{Setting} \\
\midrule
Learning Rate (SkipNets)   & 5e-4 \\
Learning Rate (Other Modules) & 1e-3 \\
Batch Size                 & 32 \\
Number of Epochs           & 100 \\
Optimizer                  & AdamW \\
Loss Function              & MSE \\
Patch Size                 & 8 \\
Width                      & 192 \\
Number of Heads            & 16 \\
Each Self-Attention Encoder's Depth              & 6 \\
MLP Ratio                  & 4 \\
Cross-Attention Encoder's Depth              & 1 \\
Image Size                 & $256\times 256$ \\
$\sigma$ (standard deviation of the Gaussian kernel)                 & 17\\
\bottomrule
\end{tabular}
\label{t3}
\end{table}
\vspace{-2em}
\subsection{Dataset and Training}
\begin{figure}
    \centering
    \includegraphics[width=0.8\linewidth]{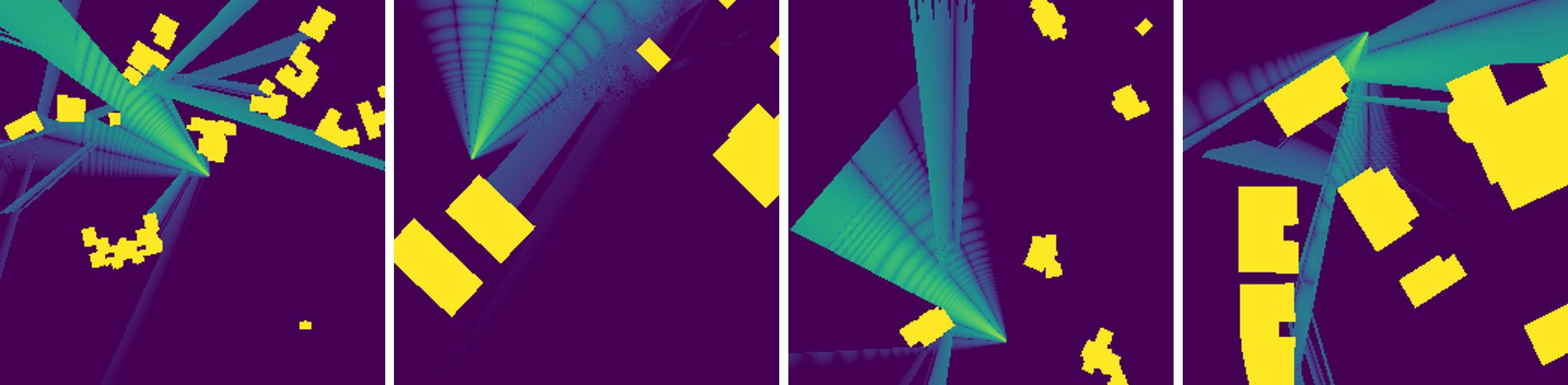}
    \caption{Visualization of the dataset}
    \vspace{-1em}
    \label{fig:vis}
\end{figure}
\begin{table*}[t]
\centering\footnotesize
\caption{Comparison across three metrics. All baselines were reproduced using their published implementations and the
hyperparameters recommended in the original papers to ensure fairness. \textbf{Localization Error} and \textbf{RMSE}: lower is better; \textbf{SSIM}: higher is better.
Best per row and in \textbf{bold}, and SOTA baselines are also highlighted in \textbf{bold} as reference for relative improvements. For Localization Error, SymNet shows \% improvement vs. \textbf{DSLoc};
for \textbf{RMSE} and \textbf{SSIM}, SymNet shows \% improvement vs. \textbf{ViT-RefineNet}.}
\begin{adjustbox}{width=\textwidth}
\begin{tabular}{
l
c c c c c  
c c c c c  
c c c c c  
}
\toprule
\multirow{2}{*}{\makecell{\# samples\\(random)}} &
\multicolumn{5}{c}{\textbf{Localization Error (m)}} &
\multicolumn{5}{c}{\textbf{RMSE}} &
\multicolumn{5}{c}{\textbf{SSIM}} \\
\cmidrule(lr){2-6}\cmidrule(lr){7-11}\cmidrule(lr){12-16}
& \makecell{\textbf{SymNet}\\\scriptsize($\Delta$ vs DSLoc)}
& DC\text{-}Net & \textbf{DSLoc} & TL;DL & LocNet
& \makecell{\textbf{SymNet}\\\scriptsize($\Delta$ vs ViT)}
& \textbf{ViT-RefineNet} & DC\text{-}Net & GAN\text{-}CRME & RadioUNet
& \makecell{\textbf{SymNet}\\\scriptsize($\Delta$ vs ViT)}
& \textbf{ViT-RefineNet} & DC\text{-}Net & GAN\text{-}CRME & RadioUNet \\
\midrule
20  & \textbf{27.891} \;\downg{24.73\%} & 59.705 & 37.056 & 69.779 & 61.316
     & \textbf{0.082819} \;\downg{0.38\%} & 0.083132 & 0.088771 & 0.091769 & 0.103657
     & \textbf{0.742213} \;\upg{7.00\%}  & 0.693653 & 0.670583 & 0.617352 & 0.715857 \\
40  & \textbf{14.142} \;\downg{33.96\%} & 45.051 & 21.416 & 64.182 & 39.912
     & \textbf{0.062740} \;\downg{3.82\%} & 0.065234 & 0.071234 & 0.070210 & 0.086102
     & \textbf{0.803986} \;\upg{5.10\%} & 0.764947 & 0.753524 & 0.733005 & 0.767339 \\
60  & \textbf{9.635}  \;\downg{36.86\%} & 37.503 & 15.259 & 62.899 & 29.779
     & \textbf{0.054712} \;\downg{4.95\%} & 0.057563 & 0.063032 & 0.061899 & 0.077658
     & \textbf{0.829305} \;\upg{4.23\%} & 0.795613 & 0.789412 & 0.773503 & 0.790645 \\
80  & \textbf{7.538}  \;\downg{38.01\%} & 33.124 & 12.159 & 63.079 & 24.183
     & \textbf{0.050327} \;\downg{5.47\%} & 0.053239 & 0.058185 & 0.057349 & 0.072550
     & \textbf{0.843697} \;\upg{3.74\%} & 0.813296 & 0.810267 & 0.795111 & 0.804493 \\
100 & \textbf{6.367}  \;\downg{38.22\%} & 30.144 & 10.306 & 63.688 & 20.643
     & \textbf{0.047480} \;\downg{5.80\%} & 0.050402 & 0.054882 & 0.054414 & 0.069025
     & \textbf{0.853431} \;\upg{3.41\%} & 0.825285 & 0.824221 & 0.809011 & 0.814019 \\
\bottomrule
\end{tabular}
\end{adjustbox}
\vspace{-1em}
\label{tab:all_metrics}
\end{table*}
We used the dataset provided in ViT-RefineNet paper \cite{ViTRefineNet} to train and evaluate our model. Each propagation map of the dataset is paired with multiple sampling maps to simulate different sensing conditions:
\begin{itemize}
    \item \textbf{Training set:} Each map (27,600 in total) is paired with 10 sampling maps, totaling 276,000 training samples.
    \item \textbf{Validation set:} Each map (5,900 in total) is paired with 40 sampling maps, totaling 236,000 validation samples.
    \item \textbf{Test set:} Each map (5,900 in total) is paired with 100 sampling maps.
\end{itemize}
In the dataset, the maximal possible signal is \( S_{\text{max}} = -4.5889\,\text{dB} \), and and a cutoff threshold of \( S_{\text{min}} = -120\,\text{dB} \) is applied, indicating that any signal strength below this threshold is treated as zero. Min--Max normalization is applied to the raw data and as follows:
\[
\frac{S_{\text{true}} - S_{\text{min}}}{S_{\text{max}} - S_{\text{min}}}.
\]
For each propagation map, there is a corresponding antenna map, where 1 indicates transmitter pixels and 0 indicates non-transmitter pixels. We generate the ground truth heatmap for the localization head by convolving the binary antenna map with a Gaussian kernel ($\sigma=17$) and normalizing the result to $[0,1]$:
\[\scalebox{0.8}{$
A(x,y) \;=\; 
\frac{ (\delta * G_{\sigma})(x,y) }
     { \max_{x,y} (\delta * G_{\sigma})(x,y) },
G_{\sigma}(u,v) = \frac{1}{2\pi\sigma^{2}}
\exp\!\left(-\frac{u^{2}+v^{2}}{2\sigma^{2}}\right).$}
\]
where $\delta(x,y)$ is the binary indicator of antenna locations, and $G_{\sigma}$ is the two-dimensional Gaussian kernel. 

The training settings are listed in Table \ref{t3}. See Figure \ref{fig:vis} for the visualization of the dataset.

\subsection{Metric}
To study the strength and shape of the signal propagation, we use \textbf{RMSE} to quantify intensity error and \textbf{SSIM} \cite{ssim} to assess perceptual quality. \textbf{SSIM} is computed following the standard definition \cite{ssim}, using local Gaussian-weighted windows averaged over the image. Specifically,
\vspace{-0.3em}
\[\scalebox{0.8}{$
\textbf{RMSE}=\sqrt{\tfrac{1}{N}\sum_{i=1}^N (y_i-\hat{y}_i)^2},\quad
\textbf{SSIM}(x,y)=\frac{(2\mu_x\mu_y+C_1)(2\sigma_{xy}+C_2)}
{(\mu_x^2+\mu_y^2+C_1)(\sigma_x^2+\sigma_y^2+C_2)},$}
\]
where $N$ is the number of non-building pixels, $y_i$ is the predicted normalized signal value, and $\hat{y}_i$ is the normalized ground-truth signal value; where $\mu_x,\mu_y$ are the mean values of $x$ and $y$, 
$\sigma_x^2,\sigma_y^2$ are the variances, 
$\sigma_{xy}$ is the covariance, 
and $C_1,C_2$ are small constants to stabilize the division.
 To measure the accuracy of transmitter localization, we use Euclidean distance from predicted location to ground truth location. We take the centroid of the prediction of the localization head as the predicted transmitter location. 

\subsection{Impact of Number of Sampled Points (RQ1)}

See Table \ref{tab:all_metrics}. From 20 to 100 sampled points, \textbf{SymNet} remains the best on localization/pixel-wise errors and is competitive or superior on structural similarity; moreover, its advantages on error-sensitive metrics \emph{increase} with sampling density.

\textbf{Localization error} becomes much lower.
Across 20/40/60/80/100 points, SymNet reduces error by \textbf{34.36\%} on average relative to DSLoc. The margin grows steadily with more samples, indicating stronger sample efficiency.

\textbf{Pixel-wise error} of heatmap prediction is also lower. 
SymNet improves over ViT-RefineNet by \textbf{4.08\%} on average. The improvement is monotonic with sampling density, suggesting better performance as more information becomes available. It also improves over the omnidirectional SOTA DC-Net by 15\% on average, demonstrating that our design is better suited to directional propagation scenarios where angular sensitivity and multipath effects dominate.

\textbf{Structural similarity} of heatmap prediction becomes higher.
Compared with ViT-RefineNet, SymNet achieves average \textbf{+4.70\%} \textbf{SSIM}. 

In summary, at low sampling (20 points), \textbf{SymNet} achieves the largest \emph{absolute} localization gain—about \textbf{10 m} vs.\ DSLoc, and the largest \textbf{SSIM} gains vs. ViT-RefineNet. As sampling increases, the absolute localization accuracy gap narrows, but SymNet \emph{consistently outperforms} baselines on error metrics; \textbf{RMSE} and \textbf{SSIM} improve consistently; \textbf{RMSE} decreases more markedly, while \textbf{SSIM} increases less as the number of sampled points grows. We therefore emphasize the strong low-sample advantage and consistent superiority. \textit{Crucially, DC-Net is an \emph{omnidirectional} SOTA; when transferred to our \emph{directional} setting, both localization and heatmap generation degrade substantially compared with our model} (e.g., 20 pts: 59.70\,m vs.\ 27.89\,m; 100 pts: 30.14\,m vs.\ 6.37\,m—\(\sim\)2.1\(\times\)–4.7\(\times\) higher error), and it also trails on pixel-wise error and \textbf{SSIM} (e.g., 20 pts \textbf{RMSE} 0.0888 vs.\ 0.0828; 100 pts \textbf{SSIM} 0.8277 vs.\ 0.8534). This highlights a \emph{task mismatch}: architectures tuned for omnidirectional propagation (which can reach sub-1\,m on their native task) transfer poorly to directional settings.

\begin{table}[t]
\centering\footnotesize
\caption{Performance by positive-ratio bins. Localization Error \&  \textbf{RMSE}: lower is better;  \textbf{SSIM}: higher is better. \# of samples $\in [20, 100]$. ViT-R.N. refers to ViT-RefineNet \cite{ViTRefineNet}.}
\vspace{-1em}
\begin{adjustbox}{width=\columnwidth}
\begin{tabular}{l
c c
c c
c c}
\toprule
\multirow{2}{*}{\makecell{\textbf{pr}\\ (\%)}} &
\multicolumn{2}{c}{\upg{\textbf{SSIM}}} &
\multicolumn{2}{c}{\downg{\textbf{Localization Error (m)}} } &
\multicolumn{2}{c}{\downg{\textbf{RMSE}}} \\
\cmidrule(lr){2-3}\cmidrule(lr){4-5}\cmidrule(lr){6-7}
& SymNet & ViT-R.N. & SymNet & DSLoc & SymNet & ViT-R.N. \\
\midrule
$[0, 10)$   & \textbf{0.702} \;\upg{8.56\%}  & 0.647 & \textbf{43.029} \;\downg{15.98\%} & 51.207 & 0.099 \;\upr{3.46\%} & \textbf{0.096} \\
$[10, 20)$  & \textbf{0.775} \;\upg{5.89\%}  & 0.732 & \textbf{19.810} \;\downg{32.77\%} & 29.475 & \textbf{0.072} \;\downg{2.46\%} & 0.074 \\
$[20, 30)$  & \textbf{0.806} \;\upg{5.01\%}  & 0.767 & \textbf{12.348} \;\downg{40.16\%} & 20.632 & \textbf{0.062} \;\downg{4.89\%} & 0.065 \\
$[30, 40)$  & \textbf{0.823} \;\upg{4.49\%}  & 0.787 & \textbf{9.069}  \;\downg{43.32\%} & 15.997 & \textbf{0.056} \;\downg{5.88\%} & 0.059 \\
$[40, 50)$  & \textbf{0.834} \;\upg{4.16\%}  & 0.801 & \textbf{7.255}  \;\downg{44.79\%} & 13.135 & \textbf{0.052} \;\downg{6.46\%} & 0.056 \\
$[50, 60)$  & \textbf{0.843} \;\upg{3.90\%}  & 0.811 & \textbf{6.150}  \;\downg{45.06\%} & 11.190 & \textbf{0.050} \;\downg{6.62\%} & 0.053 \\
$[60, 70)$  & \textbf{0.849} \;\upg{3.72\%}  & 0.818 & \textbf{5.432}  \;\downg{44.63\%} & 9.810  & \textbf{0.048} \;\downg{6.72\%} & 0.052 \\
$[70, 80)$  & \textbf{0.854} \;\upg{3.55\%}  & 0.824 & \textbf{4.952}  \;\downg{43.74\%} & 8.806  & \textbf{0.047} \;\downg{6.73\%} & 0.050 \\
$[80, 90)$  & \textbf{0.858} \;\upg{3.40\%}  & 0.829 & \textbf{4.625}  \;\downg{42.50\%} & 8.039  & \textbf{0.046} \;\downg{6.66\%} & 0.049 \\
$[90, 100]$ & \textbf{0.861} \;\upg{3.27\%}  & 0.834 & \textbf{4.362}  \;\downg{41.12\%} & 7.412  & \textbf{0.045} \;\downg{6.53\%} & 0.048 \\
\bottomrule
\end{tabular}
\end{adjustbox}
\vspace{-2em}
\label{tab:posratio_bins_all}
\end{table}
\begin{table*}[t]
\caption{Results of the ablation study across three metrics. No D' shows \% improvement vs. original \textbf{SymNet}.
}
\centering
\resizebox{\textwidth}{!}{
\begin{tabular}{l
c c c c c
c c c c c
c c c c c
}
\toprule
& \multicolumn{5}{c}{\downg{\textbf{RMSE}}}
& \multicolumn{5}{c}{\upg{\textbf{SSIM}}}
& \multicolumn{5}{c}{\downg{\textbf{Localization Error (m)}}} \\
\cmidrule(lr){2-6}\cmidrule(lr){7-11}\cmidrule(lr){12-16}
\# samples &
SymNet & No D' & Single-task & 2-CAB & 0-CAB &
SymNet & No D' & Single-task & 2-CAB & 0-CAB &
SymNet & No D' & Single-task & 2-CAB & 0-CAB \\
\midrule
20  &
\textbf{0.0828} & 0.0853 \;\upr{2.97\%} & 0.1639 & 0.1023 & 0.1646 &
0.742 & \textbf{0.763} \;\upg{2.82\%} & 0.0987 & 0.6535 & 0.0888 &
\textbf{27.891} & 29.503 \;\upr{5.78\%} & 91.948 & 56.625 & 91.814 \\
40  &
\textbf{0.0627} & 0.0659 \;\upr{4.96\%} & 0.1490 & 0.0908 & 0.1496 &
0.804 & \textbf{0.814} \;\upg{1.31\%} & 0.1259 & 0.6774 & 0.1149 &
\textbf{14.142} & 15.490 \;\upr{9.54\%} & 82.136 & 47.725 & 81.154 \\
60  &
\textbf{0.0547} & 0.0578 \;\upr{5.60\%} & 0.1342 & 0.0852 & 0.1352 &
0.829 & \textbf{0.836} \;\upg{0.80\%} & 0.1704 & 0.6834 & 0.1561 &
\textbf{9.635} & 10.721 \;\upr{11.27\%} & 74.767 & 43.189 & 73.486 \\
80  &
\textbf{0.0503} & 0.0533 \;\upr{5.99\%} & 0.1204 & 0.0816 & 0.1215 &
0.844 & \textbf{0.848} \;\upg{0.54\%} & 0.2423 & 0.6852 & 0.2203 &
\textbf{7.538} & 8.518 \;\upr{13.00\%} & 68.848 & 40.298 & 67.477 \\
100 &
\textbf{0.0475} & 0.0504 \;\upr{6.22\%} & 0.1082 & 0.0790 & 0.1094 &
0.853 & \textbf{0.857} \;\upg{0.37\%} & 0.3769 & 0.6855 & 0.3337 &
\textbf{6.367} & 7.317 \;\upr{14.91\%} & 62.728 & 38.319 & 61.652 \\
\bottomrule
\end{tabular}}
\label{tab:abl_combined}
\end{table*}

\begin{figure*}[t]
    \centering
    \includegraphics[width=0.85\textwidth]{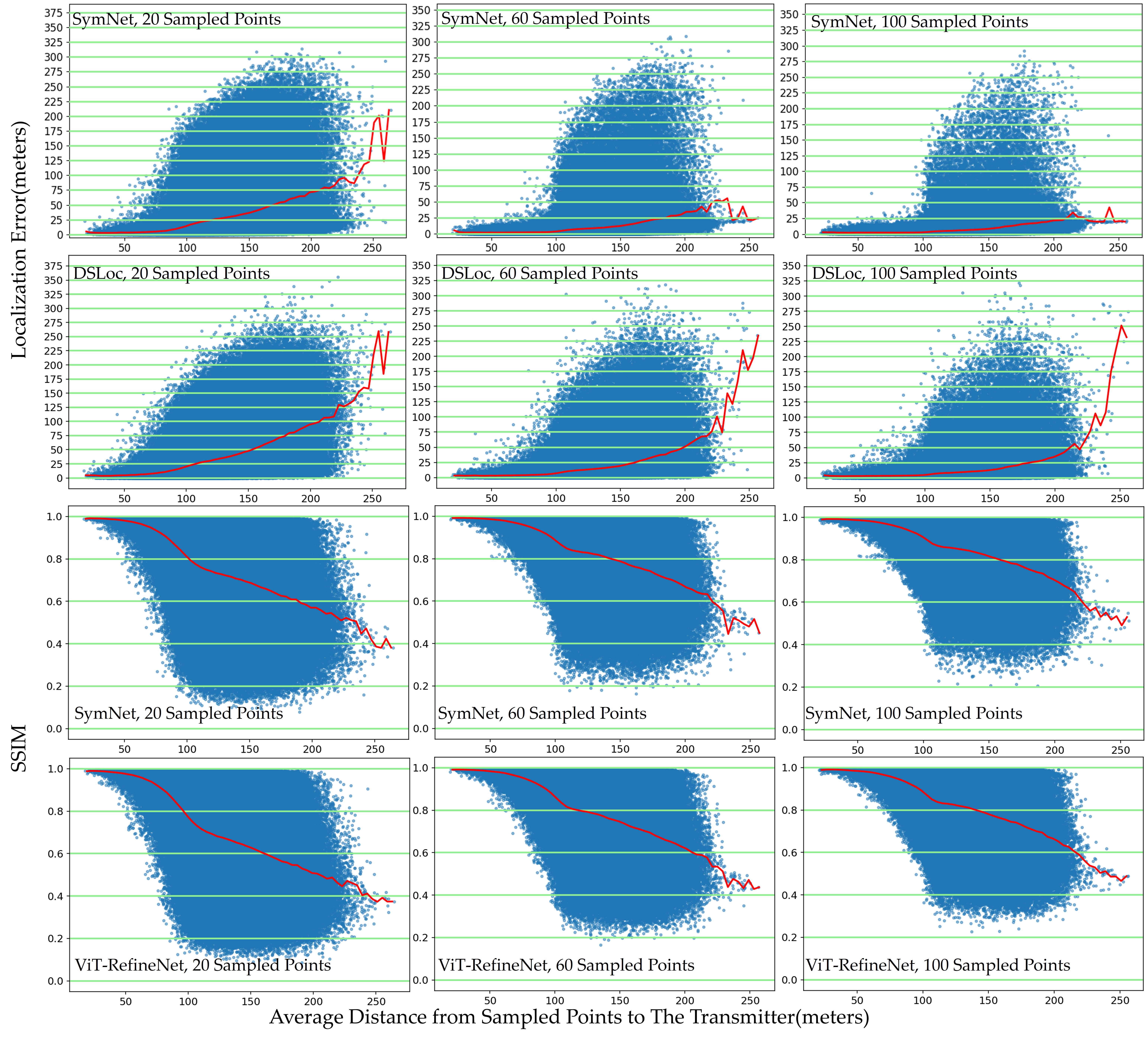}
\caption{Localization error/ \textbf{SSIM} versus the average distance from sampled points to the transmitter ($d_{\text{mean}}$) under 20 and 100 sampling points. Blue dots denote individual samples, and the red line represents the binned average Localization error/SSIM.}
    \vspace{-1.5em} 
    \label{P2}
\end{figure*}

\subsection{Impact of Positive Ratio (RQ2)}
See Table \ref{tab:posratio_bins_all}. Our model outperforms all the SOTAs in almost every scenario. In the localization task, the prediction error of our model reduces by 30\% compared to the localization SOTA DSLoc when the positive ratio is greater than 10\%; in the radio map reconstruction task, the \textbf{RMSE} of our model reduces by 5\% on average and \textbf{SSIM} increases by 5\% on average, demonstrating that our model's robustness across different positive ratios.

\subsection{Impact of Average Distance from Sampled Points to Transmitter (RQ3)}
For the localization problem, as shown in Figure \ref{P2}, our method exhibits reduced sensitivity to sampling distance. When all sampling points are located within 200 m on average, the averaged localization error remains within 75 m under 20 sampling points, while DSLoc exceeds 100 m under the same conditions. When 60 or 100 points are sampled, the average localization error is roughly bounded by 50 m, while DSLoc exceeds 75 m in the same conditions. This indicates that our approach maintains robustness in far-field scenarios where the SOTA method degrades significantly. 

For the radio map generation task, we only show how \textbf{SSIM} changes when the average distance from sampled points to the transmitter is increasing, \textbf{RMSE} follows the same pattern. The results in Figure \ref{P2} demonstrate that SymNet is more robust when increasing the transmitter distance compared to ViT-RefineNet: with only 20 sampled points, the average SSIM of SymNet remains above 0.6 until 200 m, whereas ViT-RefineNet already drops to the same level beyond 170 m. The behavior of two models in other scenarios are similar. This indicates that SymNet preserves structural similarity more effectively in far-field scenarios, highlighting its stronger generalization under sparse and distant sampling conditions. 
\subsection{\textbf{Ablation Study (RQ4)}}
We performed an ablation study to assess the contribution of each design. To align with real-world conditions, we restricted the analysis to random sampling and varied the number of sampled points. We then examined the effect of the key modifications:

\begin{itemize}
    \item {\bf No D':} removing the D' channel.
    \item {\bf 0/2 Cross-Attention Blocks(CABs):} removing all cross-attention blocks to test whether cross-task interaction is necessary, 
or doubling them with separate projection matrices to test whether the two tasks should share the same projection and whether it is better to have more cross-attention blocks. 

    \item {\bf Single-Task Path:}  removing one branch of the joint architecture and training two separate models for radio map reconstruction and localization, to examine whether joint learning is essential for convergence.
\end{itemize}
The ablation study result in Table \ref{tab:abl_combined} shows that removing the D' channel leads to a slight improvement in \textbf{SSIM} (up to +2.8\%), but at the cost of significant increases in localization error and \textbf{RMSE} (up to +15\%), indicating that D' contributes to reducing reconstruction error while preserving structural similarity. Variants altering the cross-attention design converge only to degenerate solutions with substantially worse results.
In 0-CAB, the lack of complementary guidance leads to trivial convergence, yielding behavior similar to the single-task path variant. in 2-CAB, the independent projections create inconsistent feature spaces and unstable optimization.

\section{Conclusion}
In this work, we proposed SymNet, a cross-attention–based architecture for directional radio map reconstruction and transmitter localization. Through extensive experiments, we demonstrated that SymNet achieves superior accuracy and robustness compared to state-of-the-art baselines such as ViT-RefineNet. Notably, SymNet remains less sensitive to the distance of sampling points: even when the average distance of sampled points exceeds 100 m, the model maintains stable localization errors and structural similarity, whereas competing methods degrade sharply.

Our ablation study confirmed the importance of the joint design. Removing the D' channel results in a marginal \textbf{SSIM} gain but simultaneously increases localization error and \textbf{RMSE} by a considerable margin, underscoring its role in balancing structural fidelity and numerical precision. 

Other simplified variants, including modifications to the cross-attention block, separated single-task training, or alternative head designs, converged to ineffective solutions under the same training conditions. This indicates that designs are essential for effective learning.

Overall, these findings highlight a previously overlooked challenge in wireless sensing: in directional propagation, the strongest signal often deviates from the transmitter location, making radio map reconstruction and localization complementary rather than interchangeable tasks. SymNet preserves both fine-grained structural information and global localization accuracy under sparse and distant sampling conditions, providing a robust and scalable solution for real-world deployment. Future work will extend SymNet to multi-transmitter scenarios and investigate integration with physical propagation priors to further enhance generalization.
\section{Acknowledgment}
Research was partially sponsored by the Army Research Labora-
tory and was accomplished under Cooperative Agreement Number
W911NF-23-2-0014. The views and conclusions contained in this
document are those of the authors and should not be interpreted
as representing the official policies, either expressed or implied,
of the Army Research Laboratory or the U.S. Government. The
U.S. Government is authorized to reproduce and distribute reprints
for Government purposes notwithstanding any copyright notation
herein.

{
    \small
    \bibliographystyle{ieeenat_fullname}
    \bibliography{main}
}

\end{document}